\documentclass[twocolumn]{aastex701}

\usepackage{amsmath,amssymb}
\usepackage{booktabs}
\usepackage{enumitem}
\usepackage{graphicx}
\usepackage{url}

\hypersetup{hypertexnames=false}

\shorttitle{Accelerating Chemical Kinetics for Exoplanet Atmospheres using Neural Networks}
\shortauthors{Malsky et al.}
\submitjournal{ApJ}

\begin{document}

\title{Accelerating Chemical Kinetics for Exoplanet Atmospheres using Neural Networks}

\author{Isaac Malsky}
\affiliation{Jet Propulsion Laboratory, California Institute of Technology, Pasadena, CA 91109, USA}
\affiliation{Department of Earth and Planetary Sciences, University of California Santa Cruz, Santa Cruz, CA 95064, USA}
\email[show]{isaacmalsky@gmail.com}

\author{Xi Zhang}
\affiliation{Department of Earth and Planetary Sciences, University of California Santa Cruz, Santa Cruz, CA 95064, USA}
\email{}

\author{Tiffany Kataria}
\affiliation{Jet Propulsion Laboratory, California Institute of Technology, Pasadena, CA 91109, USA}
\email{}

\author{Matthew Graham}
\affiliation{Division of Physics, Mathematics and Astronomy, California Institute of Technology, Pasadena, CA 91125, USA}
\email{}

\author{Ziyu Huang}
\affiliation{Daniel Guggenheim School of Aerospace Engineering, Georgia Institute of Technology, Atlanta, GA 30332, USA}
\email{}

\author{Boris Bonev}
\affiliation{NVIDIA Corporation, Santa Clara, CA 95051, USA}
\email{}

\author{Shang-Min Tsai}
\affiliation{Department of Earth and Planetary Sciences, University of California, Riverside, CA 92521, USA}
\affiliation{Institute of Astronomy and Astrophysics, Academia Sinica, Taipei 10617, Taiwan}
\email{}

\author{Elspeth K.H. Lee}
\affiliation{Center for Space and Habitability, University of Bern, Gesellschaftsstrasse 6, CH-3012 Bern, Switzerland}
\email{}

\correspondingauthor{Isaac Malsky}

\begin{abstract}
Observations increasingly reveal the coupled radiative, chemical, and dynamical processes that shape exoplanet atmospheres. Interpreting these atmospheres requires models that can capture this complexity. However, multidimensional models remain fundamentally limited by computational cost, and answering key questions requires simulating the governing physical mechanisms at speeds classical methods cannot achieve. As a result, models often rely on simplifying approximations, such as equilibrium chemistry, even when those assumptions miss important effects. There is a pressing need for fast and accurate chemical kinetics solvers to model planetary atmospheres. Here we present a machine learning local-box chemical kinetics solver for exoplanet atmospheres using a residual flow-map architecture. We demonstrate that this surrogate model is several orders of magnitude faster than a classical solver, achieving microsecond-scale inference while retaining percent-level accuracy. The surrogate model covers a parameter space that spans $T=300$--$3000$~K, $P=10^{-6}$--$10^{4}$~bar, $\Delta t=10^{-3}$--$10^{8}$~s, and compositions ranging from $10^{-2}$ to $10^{3}$ times solar in both C/O ratio and metallicity. Our model outperforms several commonly used machine learning architectures and performs robustly under the extreme stiffness characteristic of atmospheric chemistry. The machine learning framework presented here is a flexible and efficient approach to emulating state-to-state flow-map problems that commonly arise in numerical simulations.
\end{abstract}

\keywords{Machine Learning --- Chemical Kinetics --- Computational Chemistry}

\section{Introduction}
Chemistry plays a crucial role in governing the structure of exoplanet atmospheres by setting molecular abundances and radiative properties \citep{Heng2017}. The physics is incredibly complex: thousands of reaction pathways compete to set atmospheric abundances, mediated by transport, photolysis, condensation, and other processes \citep[e.g.,][]{moses2011, Tsai2021}. The characteristic timescales of these processes vary many orders of magnitude based on the local pressure and temperature. Although the atmosphere may be dominated by a few key species, less abundant species can affect atmospheric opacity and produce observable spectral features. With the advent of JWST---and increasingly high-fidelity observations---models that can accurately capture atmospheric chemistry are critical to characterize exoplanet atmospheres.

Three-dimensional climate models, known as general circulation models (GCMs), simulate the spatial inhomogeneities, physical processes, and dynamical effects necessary to interpret inherently multidimensional exoplanet observations \citep[e.g.,][]{Showman2009, Rauscher2010, Mendonca2018}. Broadly, these simulations operate by discretizing the 3D atmosphere into grid cells and evolving the model forward in time. This allows the model to capture the three-dimensional structure of the atmosphere in latitude, longitude, and altitude.

However, multidimensional models are computationally demanding. Simulations often evolve tens of thousands of grid cells per time step, thousands of time steps per model day, and thousands of model days. These costs also depend on the spatio-temporal resolution of the simulations. Increasing vertical or horizontal resolution, or decreasing the model time step, further increases the cost and runtime. Even when an accurate method exists for modeling a physical process, as in the case of full chemical kinetics, it may be impractical to include within larger models. Ideally, exoplanet observations would be interpreted using an ensemble of 3D simulations spanning all plausible scenarios---for example, different cloud/haze compositions, atmospheric metallicities, and C/O ratios. In practice, the cost of a single simulation makes densely sampled high resolution parameter sweeps either expensive or unfeasible.

Full chemical kinetics is particularly computationally demanding \citep{yung1984, moses2011, hu2012a, Zahnle2014, Tsai2021}. Solving the chemical evolution requires calculating the time-dependent abundances of many interacting species, coupled through large reaction networks. The resulting systems are stiff: the timescales of the fastest reactions are many orders of magnitude shorter than the timescales of the slowest reactions \citep{Tsai2017}. Integrators are forced to use small time steps to remain stable and accurate. As a result, methods that are computationally feasible for standalone chemistry calculations do not scale to multidimensional models.

Chemistry has been implemented within GCMs at a range of complexities. For example, chemical relaxation methods replace full kinetic integration with relaxation toward equilibrium abundances on parameterized chemical timescales \citep[e.g.,][]{Cooper2006, Zahnle2014, Tsai2018}. Alternatively, reduced-network approaches use a smaller set of species and reactions, allowing nonequilibrium chemistry to be evolved more efficiently in multidimensional simulations. For example, mini-chem reduces the full VULCAN network to a small set of major species and net reactions, reproducing the dominant thermochemical behavior at substantially lower cost \citep{Tsai2022, Lee2023}. These methods have enabled chemistry in multidimensional models, but they still require prescribed relaxation behavior, bespoke reaction networks, or trade-offs between speed and accuracy. These simplifications can limit physical accuracy and struggle to capture how transport and kinetics jointly drive atmospheric abundances out of equilibrium \citep[e.g.,][]{moses2011, Tsai2018, Drummond2018b, Venot2019}.

Machine learning (ML) emulation offers a computationally practical way to include complex physical processes within larger models. Emulators are ML frameworks that approximate the input-to-output mapping of classical, deterministic physical solvers \citep{Brunton2022}. These models are trained on examples from the solver to reproduce its solutions efficiently. While the underlying physical process may be computationally demanding (e.g., coupled partial differential equations), machine learning emulators can serve as surrogate models, asymptoting to the accuracy of the classical solver at a fraction of the computational cost \citep{Rasmussen2006}. 

Surrogate models for atmospheric chemical kinetics were first developed for Earth simulations. Early work showed that neural networks could emulate stiff 0D atmospheric chemistry and multiphase aerosol chemistry, demonstrating that ML surrogates can reproduce chemically complex time evolution at reduced computational cost \citep{Huang2022, Berkemeier2023}. More recent approaches have extended these ideas, with the goal of capturing particularly stiff systems \citep[e.g.,][]{Goswami2024, Nockolds2025}. In exoplanet atmospheres, \citet{Hendrix2023} used a long short-term memory (LSTM) architecture \citep{Hochreiter1997} to accelerate 1D disequilibrium chemistry calculations. More recently, \citet{Vojtekova2025} introduced CHEXANET, a U-Net model \citep{Ronneberger2015} for chemistry in hot-Jupiter atmospheres. These studies demonstrate that ML can accelerate chemical kinetics, but they are generally tied to specific networks, regimes, or profile-level predictions rather than a local state-to-state solver for arbitrary atmospheric grid cells. No existing method provides local, state-to-state chemical evolution across the broad pressures, temperatures, compositions, and time steps needed for multidimensional exoplanet simulations.

To address these issues, we create an ML model trained on the results of the VULCAN chemical kinetics code \citep{Tsai2017, Tsai2021}. Methodology regarding the training data, model architecture, and other results are detailed below. Our aim is to create a surrogate model for chemical kinetics, with the following requirements and metrics for evaluation:
\begin{enumerate}[wide=0pt, labelsep=1em, itemsep=0.5ex, label=\roman*.]
    \item \textbf{State-to-State Predictions:} The model must predict the future system state (the evolved state after some $\Delta t$) from the initial system state.
    \item \textbf{Time Step Flexibility:} The model must be able to predict state evolution for a continuous range of times (within the training data range).
    \item \textbf{Accuracy:} The model must have low errors, both on average and for worst-case predictions.
    \item \textbf{Inference Time:} The computational cost per prediction must be on the order of microseconds when amortized over a batch (a group of samples processed simultaneously) of $\mathcal{O}(100)$ samples. This reduces the computational cost to be similar to that of dynamical and radiative processes.
    \item \textbf{Parameter Space:} The model must be applicable for a range of temperatures, pressures, metallicities, and C/O ratios.
\end{enumerate}

Our model is local: for each grid cell treated as a zero-dimensional box, it determines how the mixing ratios of the tracked species evolve in time. As such, the emulator can serve as a surrogate for the core solver in atmospheric-chemistry modules that rely on a 0D kinetics scheme that evolves local mixing ratios. The model solves chemical kinetics in a single jump. Given an initial state and a single time step, it predicts the resulting species mixing ratios (with pressure and temperature held fixed). 

This article is structured as follows. In \hyperref[sec:methods]{Methods}, we describe how we created and processed the training data, define the model architecture, and summarize the training procedure and loss function. In \hyperref[sec:results]{Results}, we evaluate the model accuracy and speed, compare architectures, and document the parameter-space coverage. In \hyperref[sec:conclusions]{Discussion and Conclusions}, we summarize the main results and discuss future applications.

\section{Methods}\label{sec:methods}

\subsection{Data Generation}\label{sec:data}

We modeled the time-dependent chemical evolution of planetary atmospheres using the VULCAN chemical kinetics code \citep{Tsai2017, Tsai2021}. The simulations were conducted for a zero-dimensional (single grid cell) configuration, defined by specific pressure and temperature conditions. The chemical network comprised 52 C-H-O-N bearing species and about 1200 reactions, utilizing a fully reversed thermochemical network\footnote{\url{https://github.com/shami-EEG/VULCAN/tree/master/thermo}} without photochemistry or radiative transfer treatments. To construct a comprehensive and homogeneous dataset, we randomly sampled the initial conditions and physical parameters. We selected a primary set of twelve abundant species (H$_2$, H$_2$O, CO, CO$_2$, CH$_4$, C$_2$H$_2$, NH$_3$, N$_2$, HCN, H, OH, and O) following reduced network architectures such as mini-chem \citep{Tsai2022}. The sampling strategy prioritized hydrogen-dominated atmospheres, but also allowed for the exploration of hydrogen-poor regimes. The starting mixing ratios were initialized by sampling helium log-uniformly between $10^{-1.5}$ and $10^{-0.5}$, and the remaining trace species log-uniformly between $10^{-25}$ and $10^{0}$. Species outside this primary list were initialized with a negligible floor value of $10^{-30}$. Following initialization, all volume mixing ratios were normalized to unity. The environmental parameters were sampled independently, with temperatures drawn uniformly from the range $T \in [300, 3000]$~K and pressures sampled log-uniformly from $P \in [10^{-6}, 10^4]$~bar.

For each VULCAN simulation, the system was evolved from the randomized initial state. VULCAN employs a Rosenbrock solver with adaptive timestep control. We therefore interpolated the results onto a standardized training time grid of 100 time steps. This temporal grid included the initial condition ($t=0$) and 99 logarithmic steps ranging from $10^{-3}$ to $10^8$ seconds. 

Although the full kinetic network tracks 52 species, we reduced the output data to the twelve primary species listed above. We performed a preliminary abundance distribution analysis to confirm that these species dominate the total abundance across the majority of the parameter space. Although rare, extremely hydrogen-poor conditions can yield complex carbon-rich species; these cases represent a statistically negligible fraction of the dataset.

For model development, we used 6.2 $\times$ 10$^{7}$ independent evolutionary tracks, each containing 100 temporal snapshots\footnote{Initially, we generated $8.56 \times 10^{7}$ trajectories but discarded profiles where any species abundance falls below $10^{-30}$, as these often contained unphysical numerical instabilities.}. We split them 80\%/10\%/10\% into training, validation, and test sets. Each full trajectory can be broken up into  ${\frac{1}{2}}\, {n}({n}-1)$ training pairs, where \(n\) is the number of time points. By leveraging the parallel architecture of NASA High-End Computing resources, the calculations were completed within approximately 24 hours, producing a final dataset size of several hundred gigabytes.

\subsection{Data Normalization}

Before training, we normalize all features and targets based on only the training set data. For each chemical species, we apply a $\log_{10}$ transform (with a small positive floor of $\epsilon$=1$\times$10$^{-30}$), and then apply a z-scaling (subtracting the training-set mean and dividing by the training-set standard deviation so that each feature has zero mean and unit variance). Temperatures (which span a smaller range) are z-scaled, while pressures and $\Delta t$ values are $\log_{10}$-transformed and then linearly scaled to the interval $[0,1]$. The $\Delta t$ values are calculated from the absolute time grid of the species evolution trajectories. In our dataset, every sample has an identical time grid.

\subsection{Dataset Construction}

We sample \(N_{\mathrm{anchor}}=8\) points for each trajectory; each anchor serves as the initial state of a training pair. Of these anchors, the first is always the first time point (\(t=t_0\)) of the trajectory, while the rest are randomly sampled from the trajectory time grid. Because our dataset was generated using randomized initial abundances, the initial system states have greater diversity, and our model can more efficiently learn the underlying system dynamics by emphasizing these points.

We also randomly choose \(N_{K}=4\) target points after each anchor point to form the corresponding final state in the pair, with \(\Delta t=t_j-t_i\) and \(j>i\). Offsets are sampled with replacement, so duplicates can occur. However, because we sample a relatively small portion of the possible pairs, this is not a major inefficiency. This sampling method can under-sample the longest \(\Delta t\) values, as there are relatively few large $\Delta t$ values compared to shorter time jumps. However, forcing the first anchor point to be the initial state and sampling additional anchors across the grid provide sufficient coverage in practice. Although \(N_{K}\) or \(N_{\mathrm{anchor}}\) can be increased, this increases training cost, and we find these values are sufficient for the level of accuracy desired.

\subsection{Model Architecture}\label{sec:architecture}
\begin{figure*}
\centering
\includegraphics[width=1.0\linewidth]{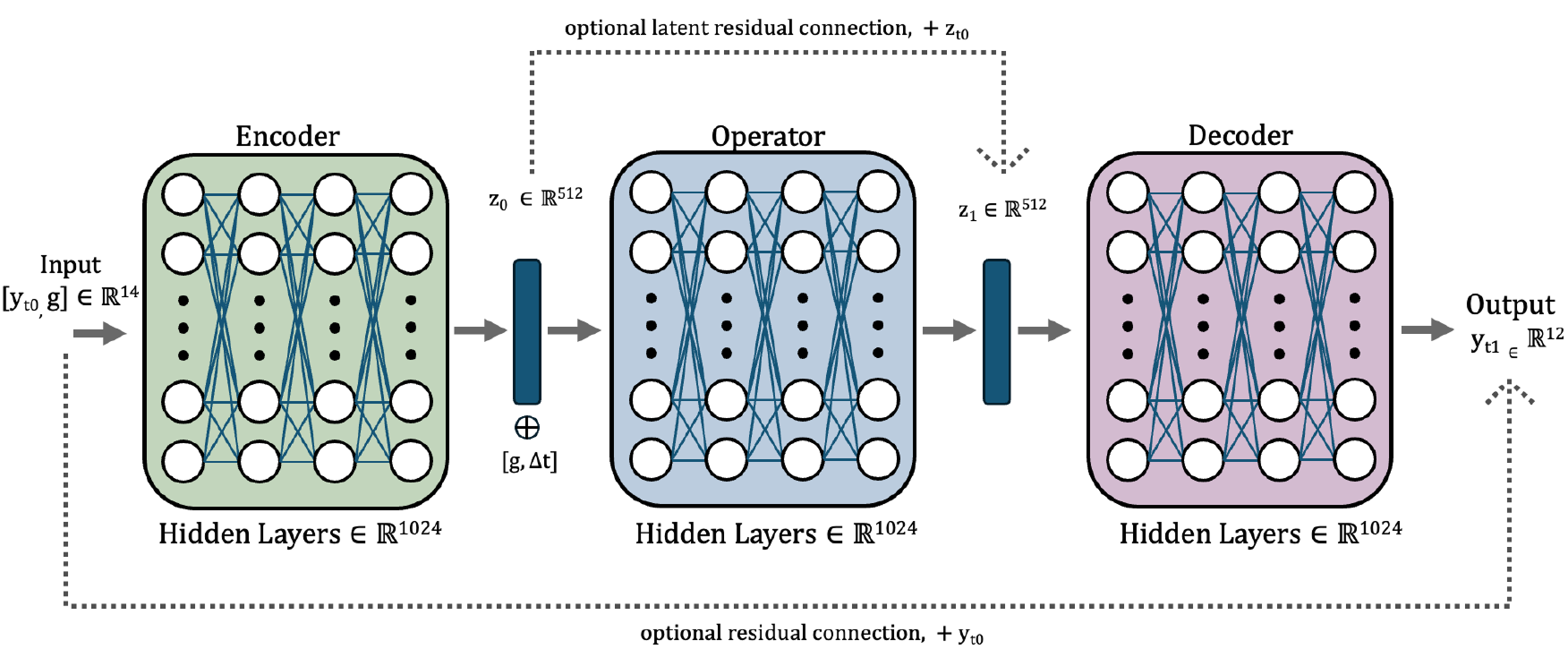}
\caption{Schematic of the flow-map encoder–operator–decoder architecture used to emulate chemical kinetics. The encoder maps the current normalized input state (and global parameters) to a latent space. The latent dynamics operator evolves this latent state using the time step and global parameters, and the decoder reconstructs the predicted future normalized species state. Optional residual connections can be applied in latent space and/or on the normalized model outputs.}
\label{fig:1}
\end{figure*}

We use an encoder--decoder architecture similar to that of an autoencoder \citep{Hinton2006}. Autoencoders were originally designed to take an input, create an intermediate representation, and reconstruct the input. Similar architectures can also be used as emulators that learn the evolution of nonlinear systems \citep[e.g.,][]{Rumel1986, Hornik1989, Lusch2018, Nockolds2025}. We refer to the model as a flow-map architecture \citep{HirschSmaleDevaney2013, ChaosBook, QIN2019, Churchill2023}. Here, we use a flow map to mean a function that advances the system state forward by a chosen time. The solution operator is approximated by

\begin{equation}
\mathbf{y}_{t+\Delta t} = \Phi(\mathbf{y}_t,\mathbf{g},\Delta t),
\end{equation}

\noindent where $\Phi$ is the flow map that advances the state $\mathbf{y}_t$ forward by a time step $\Delta t$, $\mathbf{y}_t \in \mathbb{R}^{N_s}$ encodes the (normalized) chemical species state, and $\mathbf{g}\in\mathbb{R}^{N_g}$ denotes the global parameters. A diagram of the model architecture is shown in Figure \ref{fig:1}. The model is trained as a regression model to act as a surrogate for the full VULCAN calculations. It takes the input state and advances it in time by $\Delta t$, where $\Delta t$ can take any value within the range encountered in the training set. The global parameters ($\mathbf{g}$) consist of pressure and temperature. In our case, ``global'' refers to pressure and temperature being held fixed during a single time step.

\begin{table}
    \centering
    \footnotesize
    \setlength{\tabcolsep}{3pt}
    \begin{tabular}{@{}l l@{}}
        \toprule
        \textbf{Property} & \textbf{Value} \\
        \midrule

        \multicolumn{2}{l}{\textbf{Inputs and Outputs}} \\
        Global inputs & \({N_g=2}\) (Pressure, Temperature) \\
        Species predicted & \({N_s} = 12\) (e.g., \({H_2}\)) \\
        Time input & \({\Delta t}\) \\

        \addlinespace
        \multicolumn{2}{l}{\textbf{Architecture}} \\
        Latent dimension & \({N_z=512}\) \\
        Encoder hidden sizes & [1024, 1024, 1024, 1024] \\
        Dynamics hidden sizes & [1024, 1024, 1024, 1024] \\
        Decoder hidden sizes & [1024, 1024, 1024, 1024] \\
        Activation & SiLU \\
        Residual connections & latent residual + output residual\\

        \addlinespace
        \multicolumn{2}{l}{\textbf{Training setup}} \\
        Epochs & 300 \\
        Batch size & 2048 \\
        Pairs per anchor & 4 \\
        Anchor points per trajectory & 8 \\
        Optimizer & AdamW \\
        Learning rate & \({10^{-4}}\) \\
        Minimum learning rate & \({5 \times 10^{-7}}\) \\
        Warmup epochs & 10 \\
        Weight decay & \({10^{-4}}\) \\

        \addlinespace
        \multicolumn{2}{l}{\textbf{Loss}} \\
        Objective & weighted \({\mathcal{L}_{\rm log\text{-}ratio} + \mathcal{L}_{\rm mse}}\) (Eq.~\ref{eq:loss}) \\
        Weights & (1.0, 0.5) \\

        \bottomrule
    \end{tabular}
    \caption{Model and training configuration for the chemical kinetics emulator. Each epoch takes $\sim$20 minutes on a single NVIDIA A100 80 GB PCIe.}
    \label{tab:model_specs}
\end{table}

The architecture contains three sequential building blocks: an encoder ($\mathcal{E}_{\theta_E}$) that maps the physical state to a latent representation, a latent dynamics operator ($\mathcal{F}_{\theta_F}$) that evolves this representation in time, and a decoder ($\mathcal{D}_{\theta_D}$) that reconstructs the predicted state. This design separates how the initial state is encoded/reconstructed and how it is evolved for a specified time step $\Delta t$. We can write the model symbolically as

\begin{equation}
\mathbf{y}_{t+\Delta t}
=
\mathcal{D}_{\theta_D}\!\Big(
\mathcal{F}_{\theta_F}\!\big(\mathcal{E}_{\theta_E}(\mathbf{y}_t,\mathbf{g}),\,\Delta t,\,\mathbf{g}\big)
\Big)
\label{eq:flowmap_noresiduals}
\end{equation}

Motivated by chemical evolution which is often modeled in terms of changes relative to the current state, we also create a variant of the model designed to predict the residuals rather than the full state. This can be done either in the latent space or for the output. A similar approach has been previously used \citep[e.g.,][]{He2016} and is known to aid learning for deep models. This variant for our case can be expressed as

\begin{equation}
\scalebox{0.92}{$
\mathbf{y}_{t+\Delta t}
=
\mathbf{y}_t
+
\mathcal{D}_{\theta_D}\!\Big(
\mathcal{E}_{\theta_E}(\mathbf{y}_t,\mathbf{g})
+
\mathcal{F}_{\theta_F}\!\big(\mathcal{E}_{\theta_E}(\mathbf{y}_t,\mathbf{g}),\,\Delta t,\,\mathbf{g}\big)
\Big)
$}
\label{eq:flowmap_both_residuals}
\end{equation}

\noindent where the encoded state, $\mathcal{E}_{\theta_E}(\mathbf{y}_t,\mathbf{g})$, is added to the output of the latent dynamics network, $\mathcal{F}_{\theta_F}$, and the decoded result is added to the initial state $\mathbf{y}_t$. Either residual connection can be enabled independently. Both are enabled in the model reported here. The architecture and important hyperparameters are shown in Table \ref{tab:model_specs}.

As data flows through the model, the encoder network maps the inputs $\mathbf{y}_t,\mathbf{g}\in\mathbb{R}^{N_s+N_g}$ to a latent vector $\mathbf{z}_t\in\mathbb{R}^{N_z}$. The latent dynamics network then takes $(\mathbf{z}_t, \Delta t,\mathbf{g})\in\mathbb{R}^{N_z+1+N_g}$ and outputs an evolved latent vector $\mathbf{z}_{t+\Delta t}\in\mathbb{R}^{N_z}$. Finally, the decoder network maps $\mathbf{z}_{t+\Delta t}$ back to the predicted species state $\mathbf{y}_{t+\Delta t}$. This output value is then de-normalized to physical species mixing ratios.

Our reaction network is particularly stiff. Following \citet{Tsai2017}, we define stiffness as

\begin{equation}
r = \frac{t_s}{t_f} = \frac{\max|\mathrm{Re}(\xi)|}{\min|\mathrm{Re}(\xi)|},
\label{eq:stiffness_ratio}
\end{equation}

\noindent where $t_s$ is the slowest reaction timescale and $t_f$ is the fastest, and $\xi$ are the eigenvalues of the Jacobian of the chemical kinetics system. For comparison, the Robertson system (a smaller coupled ordinary differential equation problem often used to benchmark numerical integrators; \citealt{Robertson1967}) has a stiffness ratio of $r \sim10^{10}$. However, chemical networks required for exoplanet atmospheres have $r\sim10^{20}$--$10^{30}$, implying a much more difficult numerical problem \citep{Tsai2017}. As part of this work, we compare different machine learning architectures on the Robertson problem (see Appendix~\ref{app:robertson}).

\subsection{Training}

All models are implemented and trained with \texttt{PyTorch} \citep{Paszke2019}. We train the emulator using mini-batches of sampled flow-map pairs constructed as described above, with batch size $B=2048$. Optimization is performed with AdamW (weight decay $10^{-4}$) using an initial learning rate of $10^{-4}$ \citep{Loshchilov2017}. We use a linear warmup over the first 10 epochs followed by cosine annealing to a minimum learning rate of 5 $\times 10^{-7}$. The training objective is the weighted combination of log-ratio and normalized-space MSE losses described in Eq.~\ref{eq:loss}.

Training is run on a single NVIDIA A100 80 GB PCIe GPU using automatic mixed precision in bfloat16 mode, with TF32 enabled. We enable cuDNN benchmarking and do not enforce deterministic execution. For the full VULCAN emulator configuration, training requires $\sim$20 minutes per epoch on the A100. The smaller models used for ablation tests (e.g., the models with fewer parameters shown below) do not saturate the A100, and their wall-clock time is correspondingly dominated by non-compute overheads rather than GPU throughput.

\subsection{Weighted Loss Function}
We find that a custom loss function results in improved model performance. We implement both an $L_1$ loss in $\log_{10}$ space and an MSE loss in normalized space during training:

\begin{equation}
\begin{aligned}
\mathcal{L}
&= \mathcal{L}_{\rm log\text{-}ratio} + 0.5\,\mathcal{L}_{\rm mse},\\
\mathcal{L}_{\rm log\text{-}ratio}
&= \frac{1}{N_s}\sum_{i=1}^{N_s}\left|\log_{10}\!\left(\frac{\hat a_i}{a_i}\right)\right|,\\
\mathcal{L}_{\rm mse}
&= \frac{1}{N_s}\sum_{i=1}^{N_s}\left(\hat x_i - x_i\right)^2,\\
\hat x_i
&\equiv \frac{\log_{10}\!\left(\hat a_i\right)-\mu_i}{\sigma_i},
\qquad
x_i \equiv \frac{\log_{10}\!\left(a_i\right)-\mu_i}{\sigma_i}.
\end{aligned}
\label{eq:loss}
\end{equation}

\noindent where $\mu_i$ and $\sigma_i$ are the mean and standard deviation of $\log_{10}\!\left(a_i\right)$ for species $i$, computed from the training set and used to standardize each species in log space, and $a$ represents the physical (non-normalized) species abundances. The total objective is a weighted sum of these components; specifically, we set the weights of $\mathcal{L}_{\rm log\text{-}ratio}$ and $\mathcal{L}_{\rm mse}$ to 1.0 and 0.5, respectively. Both losses are averaged over the $N_s$ predicted species.

Equation~\ref{eq:loss} is for a single target state (i.e., one initial state and the resulting state after some $\Delta t$). Although not shown here for simplicity, during training each sampled dataset item consists of one anchor state and $N_K$ target times, and we compute the loss for every target in the mini-batch. The final loss is the mean over all sampled pairs and all samples in the mini-batch (i.e., over $B \times N_K$ target states, equivalently $B \times N_K \times N_s$ scalar species terms). Other loss functions, including Huber loss and MAE, were also explored but yielded lower overall accuracy.


\section{Results}\label{sec:results}

\begin{figure*}
\centering
\includegraphics[width=1.0\linewidth]{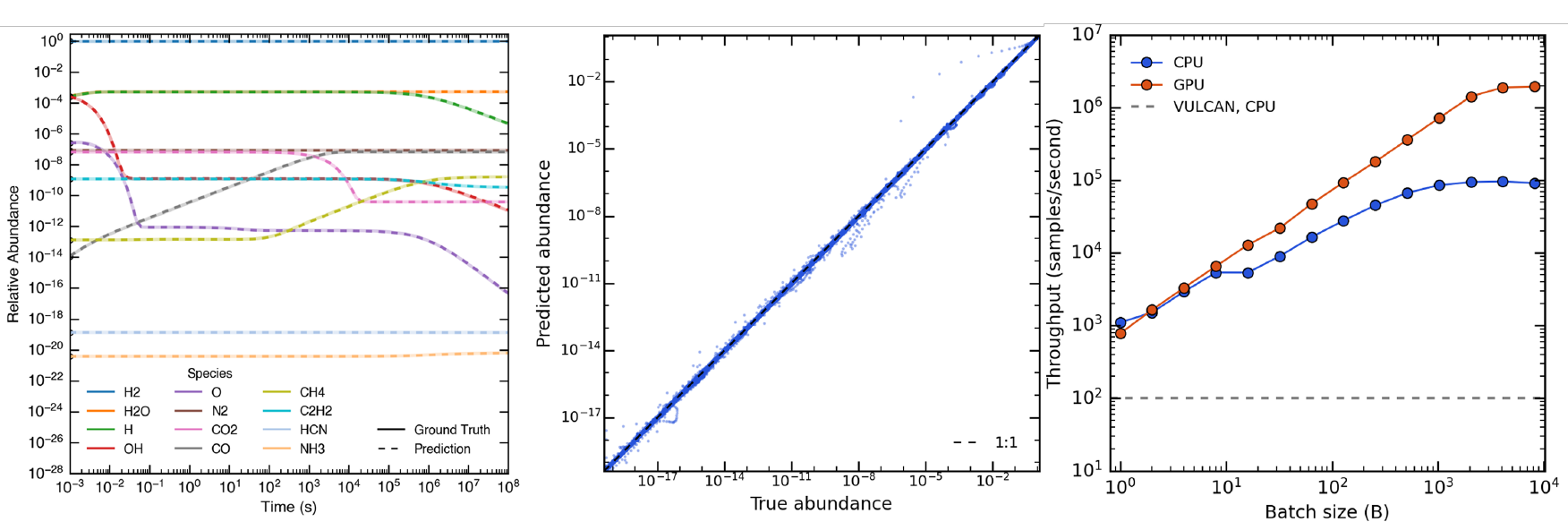}
\caption{Left: an example test-set trajectory for our trained chemical kinetics emulator (dashed lines), compared to the ground-truth VULCAN data (solid lines). For the 100 plotted times, the model takes the initial system state, the time to the final state $\Delta t$, the pressure and temperature values, and determines the final evolved state of the system. This profile was evolved at a pressure of $\sim$3.1 mbar and 2620 K. The model is able to capture the stiff dynamics for times that span more than ten orders of magnitude. Middle: the true abundance values (from VULCAN) versus predicted values for the chemical kinetics emulator. The scatter points represent $\sim$10$^{5}$ individual time predictions, for 99 predictions per trajectory, representing all 12 species predictions, all randomly chosen from the test set trajectories. Right: The inference speed of our model, run on an ARM Neoverse-V2 CPU and an A100, and amortized over batches of varying size. Inference on either a CPU or a GPU is orders of magnitude faster than the classical VULCAN chemical kinetics calculations.}
\label{fig:2}
\end{figure*}

\subsection{Accuracy}
Our machine learning model is able to accurately emulate the chemical kinetics evolution predicted by VULCAN. Figure \ref{fig:2} shows an example test-set trajectory of the model. From a single initial condition, the model predicts the temporal evolution of species abundances over times ranging from 10$^{-3}$ to 10$^{8}$ seconds. The model captures the behavior of dominant species, like H$_2$, that remain relatively stable in abundance, as well as less abundant species, like CH$_4$, that show a large dynamic range. Although the VULCAN network used to generate the training data includes 52 distinct species, the model only tracks the 12 species shown in Figure \ref{fig:2}. We track the same 12 species as \citet{Tsai2022}, as this set is the most important for governing thermal opacities. The remaining 40 species are not included as model targets or input features. This demonstrates that, for our regime, the evolution of these key species can be accurately predicted using the smaller subset of species (at least within the time domain of our non-autoregressive model). Furthermore, the model architecture is flexible with respect to the number of species. Using a smaller or larger species set requires retraining, but no changes to the underlying model---making the architecture flexible for future atmospheric studies beyond exoplanet applications.

Figure \ref{fig:2} shows the true versus predicted abundance values for $\sim$10$^{5}$ samples from the test set. We define the fractional error of a single species prediction as $|\hat a_i - a_i|/(a_i + \epsilon)$, with $\epsilon=10^{-20}$, where $a_i$ and $\hat a_i$ are the true (VULCAN) and predicted abundances of species $i$. The $\epsilon$ is added to better handle species with trace abundances. About 4.5\% of test-set targets lie below $10^{-20}$, and without an epsilon those points would skew the results while not being physically important abundances. Within these $\sim$10$^{5}$ predictions, the fractional errors corresponding to the 50th, 90th, 95th, 99th, and 99.9th percentiles were 1.7\%, 7.8\%, 11.1\%, 32.0\%, and 101.8\%. When filtered to only include true values larger than 10$^{-10}$ (dominant species), these fractional errors fall to 1.6\%, 6.9\%, 9.7\%, 23.6\%, and 95.8\%. In comparison, simplified chemistry schemes may differ from full chemical kinetics by several orders of magnitude \citep{Tsai2022}.


\subsection{Speed}
The cost of calculating each chemical kinetics prediction is on the order of microseconds per sample. This is many orders of magnitude faster than the original VULCAN calculations. At this speed, chemical kinetics calculations are no more computationally demanding than dynamical or radiative calculations within multidimensional models. This opens the door for models that can capture the interactions and feedback between these processes at unprecedented accuracy.

Figure \ref{fig:2} shows the computational cost per sample for both CPU and GPU calculations. Here, a calculation corresponds to a single $\Delta t$ jump, but includes all tracked species. For a single prediction (batch size of one), the computational cost is on the order of milliseconds. However, larger batches reduce per-sample overhead, and the cost per sample decreases approximately inversely with batch size (until plateauing for large batches). We note that climate simulations require the parallel calculation of many 0D grid cells, meaning that large-batch calculations are well suited to this type of task.

\subsection{Architectural Ablation Tests}

\begin{figure*}
\centering
\includegraphics[width=1.0\linewidth]{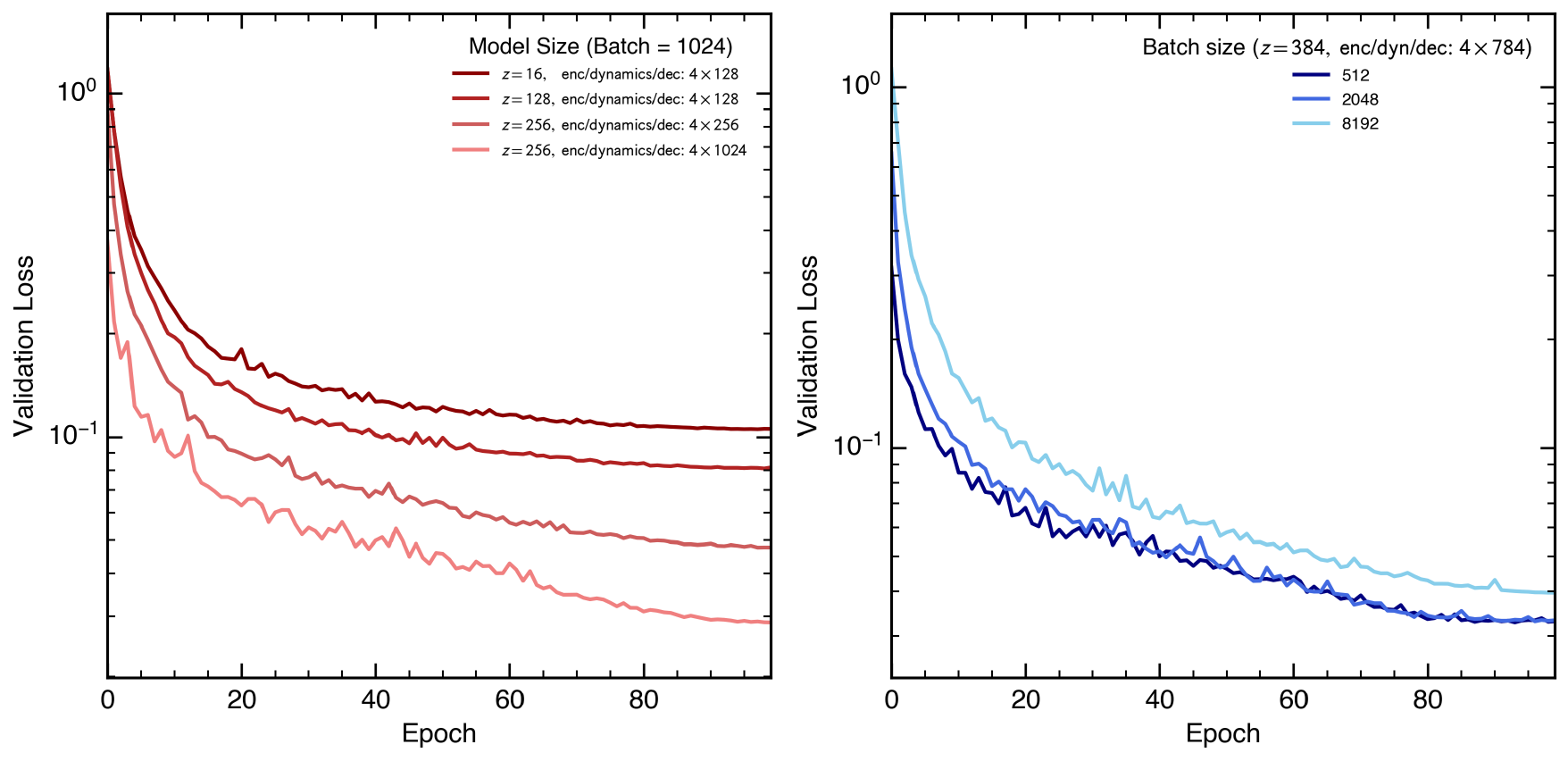}
\caption{Validation losses for models with different latent dimensionalities ($N_z$), and encoder, dynamics, and decoder network sizes (left). Larger models, and models with larger latent dimensions, perform better. Beyond a threshold, increasing batch size during training degrades model performance. Batch sizes above $\sim$2048 resulted in worse generalization (right), likely as the model had converged to a sharp minimum in the training space \citep{Keskar2016, Goyal2017}.}
\label{fig:4}
\end{figure*}

Large mini-batches can degrade generalization beyond a certain threshold \citep{Keskar2016, Goyal2017}. This has been of considerable interest given the demand for fast training on enormous datasets and the increase in available compute. Although the underlying causes remain debated, large-batch training has been associated with convergence to sharper minima of the loss landscape \citep{Keskar2016}. Batch size exploration is especially important for our problem: the model has $\sim$11.6 million parameters, but requires a large training dataset to cover the high-dimensional parameter space of chemical kinetics. We expect related works emulating physical processes to face similar constraints, e.g., networks that are small by the standards of modern deep learning (millions of parameters, rather than billions), but trained on very large datasets. Figure~\ref{fig:4} shows that our model performance drops for batches larger than $\sim$2048. In practice, we mitigate batch-size limitations by training several models simultaneously on one GPU. Additionally, we use a lower learning rate during the warm-up period at the start of training \citep{Goyal2017}.

Figure~\ref{fig:4} shows that validation loss decreases as model size increases. Both increasing the latent dimensionality and increasing the layer width decrease validation loss. For this work, we aim to balance computational cost and accuracy. Because increasing model size increases inference cost, we choose a relatively small final model that achieves multiscale accuracy with typical fractional errors at the percent level. We expect larger models to further decrease error. As noted, comparable simplified chemical kinetics schemes may have errors larger than 100\% \citep{Tsai2022}. We performed detailed parameter testing to identify an accurate set of hyperparameters. This included varying layer widths, learning rates, activation functions, weight decay, and other settings.

\subsection{Parameter Space Coverage}
\begin{figure*}
\centering
\includegraphics[width=1.0\linewidth]{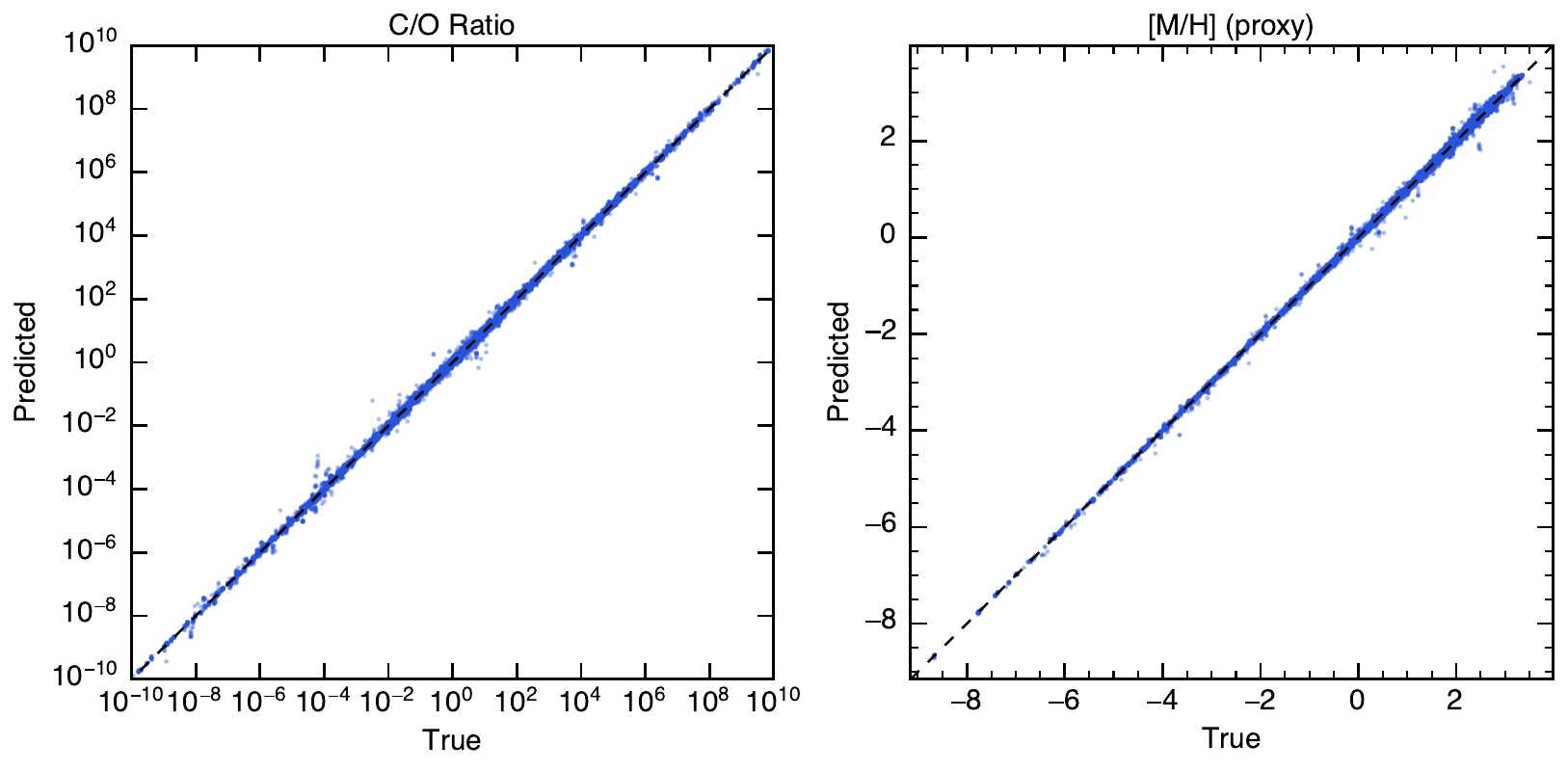}
\caption{The true versus ML-surrogate model predicted compositions: (left) C/O ratio and (right) [M/H] proxy. Because of our sampling methodology, the data span an extremely large range of compositions, and model predictions remain accurate across nearly the entire distribution. Each panel shows 20,000 hold-out test set pairs.}
\label{fig:5}
\end{figure*}

The data cover an extremely large parameter space in terms of C/O ratios and metallicity, as shown in Figure~\ref{fig:5}. Because we sample the individual species abundances log-uniformly, the initialized values span an extremely large set of mixing ratios. We note that we only track an incomplete set of metals, so the metallicity values are a proxy. Because of the broad coverage, the ML emulator is applicable across a diverse set of exoplanet atmospheres. Figure~\ref{fig:5} also shows that the accuracy of the model does not degrade across the C/O and metallicity space shown. Although extending the model presented here to other regimes would require regenerating the VULCAN training set and retraining the model, the underlying architecture and workflow would remain the same.

\subsection{mini-chem Comparisons}\label{sec:minichem}

\begin{figure*}
\centering
\includegraphics[width=1.0\linewidth]{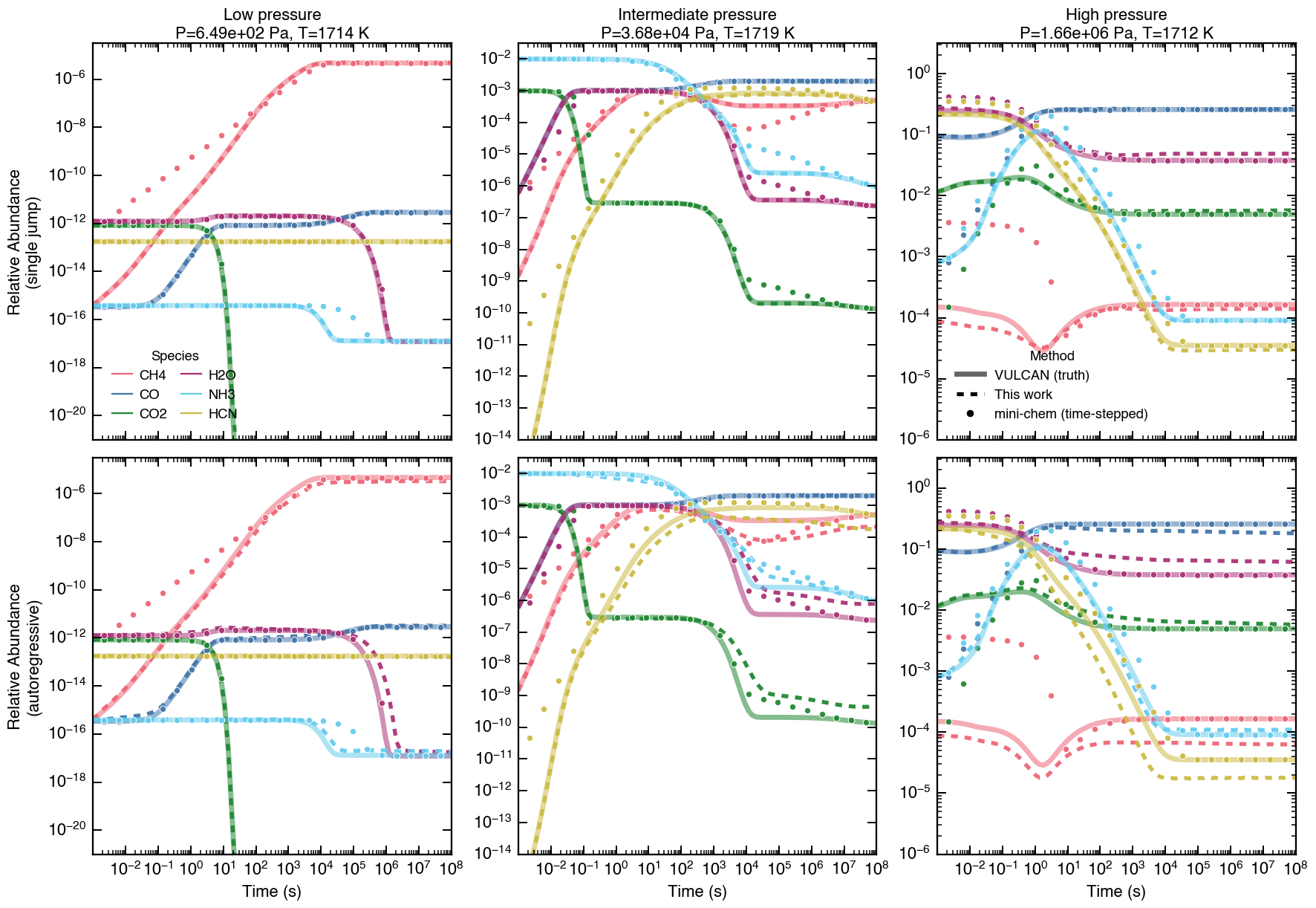}
\caption{Comparison of VULCAN, the chemical kinetics emulator presented here, and mini-chem for three representative trajectories. Solid lines show the VULCAN reference solution, dashed lines show predictions from this work, and markers show mini-chem evolved sequentially through the same time grid. The top row shows the single-jump mode of the ML emulator, and the bottom row shows the emulator run autoregressively for 99 sequential jumps.}
\label{fig:compare}
\end{figure*}

\begin{figure}
\centering
\includegraphics[width=1.0\linewidth]{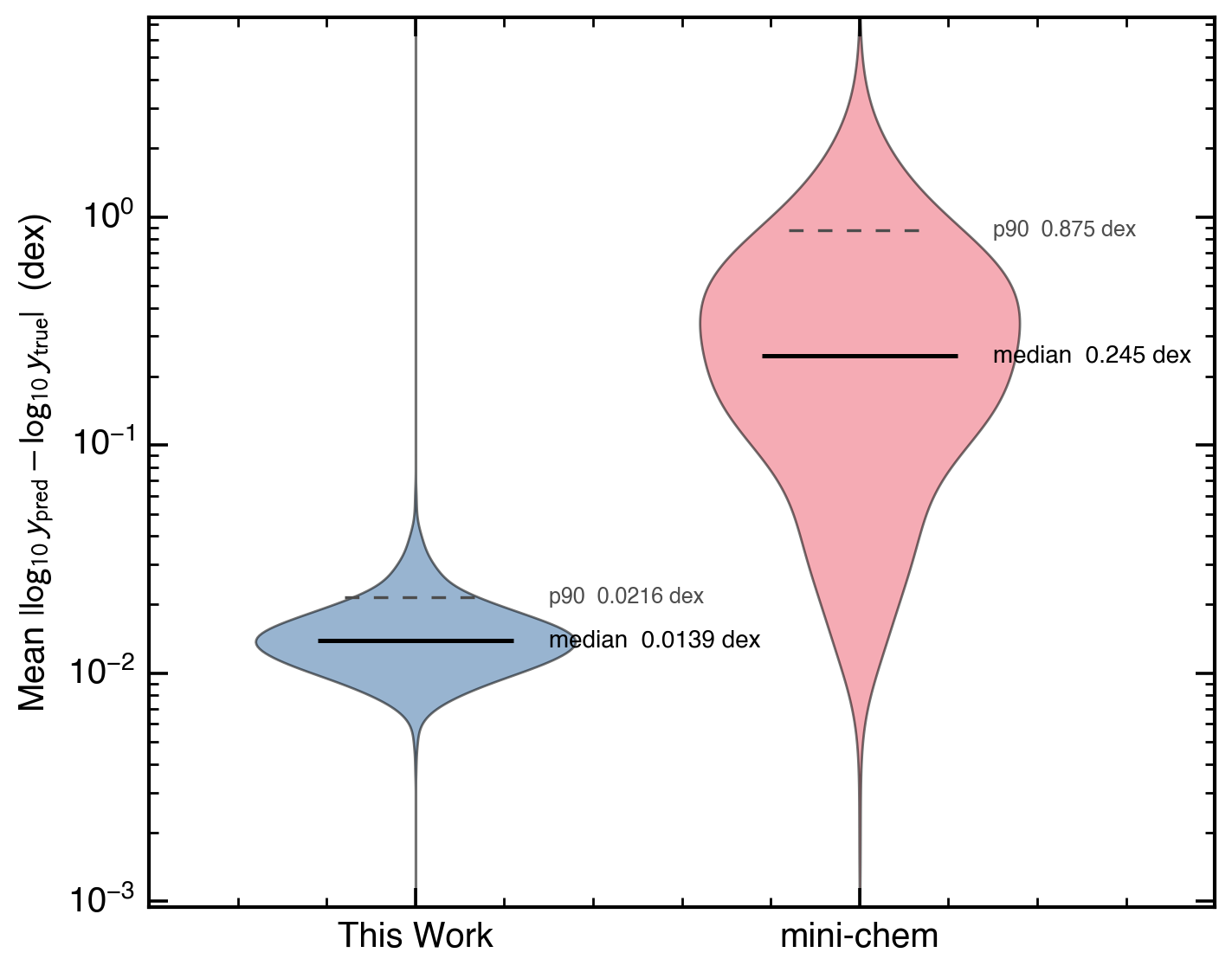}
\caption{Distribution of trajectory-averaged mean absolute $\log_{10}$ errors, relative to VULCAN, for this work and mini-chem. Errors are averaged over all 12 tracked species and all 99 output times for each trajectory, using 1000 trajectories restricted to $T = 1000$--$2500~\mathrm{K}$ and $P = 10^{2}$--$10^{7}~\mathrm{Pa}$.}
\label{fig:mini-chem_error}
\end{figure}

We compare the ML surrogate model against mini-chem\footnote{\url{https://github.com/ELeeAstro/mini_chem}}, a reduced chemical kinetics solver for gas giant exoplanet atmospheres. Mini-chem evolves a smaller set of species using T-p dependent net forward reactions calibrated against the full VULCAN network, making it fast enough for 3D GCM applications while retaining a physically interpretable reaction structure \citep{Tsai2022, Lee2023}.

Because the ML surrogate model is trained directly from a large VULCAN dataset, a single trained model can cover a broad compositional parameter space. This flexibility is one of the main advantages of the emulator: within the range covered by the training data, it does not have to be recalibrated for different thermochemical regimes. Comparatively, mini-chem is a reduced kinetic solver, so its behavior remains tied to a small set of species and physically motivated net reactions. This makes it fast and interpretable, but also means that applying mini-chem to different metallicities and C/O ratios requires generating the corresponding rates for the net reactions. 

Overall, the ML emulator shows closer agreement with the full VULCAN trajectories, although both methods capture the broad system dynamics. Figure~\ref{fig:compare} shows three example trajectories comparing mini-chem, VULCAN, and the surrogate model. We compare two modes: a single-jump mode, in which each state is predicted directly from the initial state, and an autoregressive mode, in which the model is advanced sequentially over 99 time steps. Both mini-chem and the emulator are initialized from the same VULCAN state. For clarity, we plot six representative species and three example trajectories spanning low-, intermediate-, and high-pressure conditions. Figure~\ref{fig:mini-chem_error} shows the mean absolute $\log_{10}$ error, in dex, relative to VULCAN, averaged over all 12 species and all 99 output times for each trajectory. The corresponding 90th-percentile errors are 0.0216~dex and 0.875~dex.

On an Apple M3 Pro CPU, mini-chem's cost per grid-cell calculation asymptotes to $\sim$27~$\mu$s per solver call, comparable to the ML emulator on the same hardware.

Although our model can also be run autoregressively by feeding each predicted state into the next prediction, the model used here was trained for one-shot state-to-state evolution. In autoregressive tests, the model remains accurate over short roll-outs but degrades at longer times as small prediction errors accumulate and push the input state away from the training distribution \citep{Lamb2016, Brandstetter2022}.

Autoregressive functionality is important for GCM applications, where chemical solvers must remain stable over tens of thousands of time steps \citep[e.g.,][]{Rauscher2010, Drummond2020, Steinrueck2023}. The 10th, 50th, 90th, and 95th percentile fractional errors are 0.2\%, 3.2\%, 14.5\%, and 21.1\% after ten autoregressive steps, increasing to 0.7\%, 13.2\%, 56.6\%, and 85.3\% after 99 steps. This suggests that the architecture can support autoregressive evolution, but a GCM implementation will require training procedures designed specifically to control error accumulation over long roll-outs. In future work, we will adapt training for long-horizon autoregressive tasks. In contrast, mini-chem has been validated against long VULCAN trajectories up to times of $10^{18}$~s.

\section{Discussion and Conclusions}\label{sec:conclusions}

In this work, we develop a state-to-state flow-map emulator, to predict evolving local mixing ratios within exoplanet atmospheres. The emulator is trained on VULCAN solutions, and replaces a classical time integration method. The ML model is a direct mapping from an initial state to the evolved state, rather than a time-stepped integrator. When amortized over large batches, the model presented here achieves microsecond-scale inference per sample, making chemical kinetics comparable in cost to dynamical processes within larger models, while maintaining percent-level accuracy for one-shot evolution predictions. The model covers $T \in [300, 3000]$~K, $P \in [10^{-6}, 10^4]$~bar, $\Delta t \in [10^{-3}, 10^{8}]$~s, and species abundances spanning tens of orders of magnitude. Our model also has broad composition coverage, from strongly oxygen-rich to strongly carbon-rich conditions, with C/O ratios spanning at least \(10^{-2}\) to \(10^{3}\) times solar, and from strongly subsolar to highly super-solar metallicities, spanning at least \(10^{-2}\) to \(10^{3}\) times solar.

Other architectures have been explored to model chemical kinetics. Neural ordinary differential equation models \citep{Chen2018a} learn the time evolution of a system, but advancing the state still requires numerical time integration; for stiff kinetics this can be computationally expensive. Operator-learning approaches have also been applied in related contexts \citep{Goswami2024}, including Fourier Neural Operators \citep{Li2020}. However, for our specific goal of computationally efficiently ``jumping'' from a single initial state to a final state in one step, we find that these architectures are not well suited to this problem. Lastly, Koopman models \citep{Lusch2018, Brunton2021} have been used to learn nonlinear dynamics, but in practice they enforce approximately linear evolution in a learned latent space, which can be restrictive and may not capture strongly nonlinear behavior expected for chemical kinetics \citep{Champion2019}.

Chemical kinetics occupies a high-dimensional parameter space: the evolution is sensitive to the abundances of $\mathcal{O}(10)$ species as well as pressure and temperature. The system is also extremely stiff ($r \sim 10^{20}$--$10^{30}$). Achieving percent-level accuracy required an extremely large training dataset. Our results indicate that while many machine learning architectures perform well on simpler stiff systems (such as the Robertson equations), accurate emulation of exoplanet chemical kinetics requires architectures and training frameworks tailored to stiffness and wide dynamic range. We suggest our chemical kinetics dataset can serve as a benchmark for future machine learning studies.

The residual encoder--operator--decoder architecture reproduces the classical VULCAN solutions to a median fractional error of 1.7\% per species. Although the model presented here is tailored for exoplanet chemical kinetics, the results are applicable to emulating large chemical kinetics networks. This framework can be applied more broadly in other atmospheric modeling contexts, such as emulating cloud microphysics, or other objects such as brown dwarf GCMs and sub-Neptune atmospheres which can have complex time-dependent chemistry. Through machine learning emulation techniques, one can reduce the computational cost of expensive routines, increase spatial and temporal resolutions, and enable broader parameter exploration in multidimensional models.

\section*{Data Availability}
The code for this paper is available at \url{https://github.com/imalsky/Chemulator}, and the Robertson benchmark code of Appendix~\ref{app:robertson} at \url{https://github.com/imalsky/robertson-emulator}. The trained model weights are archived in a Zenodo record: \dataset[DOI: 10.5281/zenodo.21866197]{https://doi.org/10.5281/zenodo.21866197}.



\begin{acknowledgments}
A portion of this research was carried out at the Jet Propulsion Laboratory, California Institute of Technology, under a contract with the National Aeronautics and Space Administration (80NM0018D0004). X.Z. is supported by the National Science Foundation Astronomy and Astrophysics Research Grant (AAG) 2307463, the NASA Exoplanet Research grant (XRP) 80NSSC22K0236, and the NASA Interdisciplinary Consortia for Astrobiology Research grant (ICAR) 80NSSC21K0597. Z.H. is supported by SSERVI-CLEVER NNH22ZDA020C/80NSSC23M022 and Gatech Astrobiology Fellowship. S.-M.T. is supported by the National Science and Technology Council (grant 114-2112-M-001-065-MY3). E.L. is supported by the CSH through the Bernoulli Fellowship.
\end{acknowledgments}

\appendix

\section{Robertson Benchmark}\label{app:robertson}
As a smaller proxy for the full chemical kinetics problem, we evaluate multilayer perceptron (MLP), flow-map, and DeepONet\footnote{The Deep Operator Network; from \cite{Lu2019}.} models on the Robertson problem, a coupled ODE system originally created to model an autocatalytic reaction \citep{Robertson1967}. This benchmark allows us to quantify each architecture’s prediction accuracy and to track how that accuracy degrades as stiffness is increased (by varying the reaction-rate parameters).

The Robertson problem describes the evolution of three chemical species and is widely used to benchmark numerical integrators because it yields a stiff ODE system with rate constants spanning many orders of magnitude, producing widely separated dynamical timescales \citep[e.g.,][]{HairerWanner1996}. The equations are

\begin{equation}
\begin{aligned}
\frac{dx_1}{dt} &= -p_1 x_1 + p_3 x_2 x_3, \\
\frac{dx_2}{dt} &=  p_1 x_1 - p_3 x_2 x_3 - p_2 x_2^{2}, \\
\frac{dx_3}{dt} &=  p_2 x_2^{2}.
\end{aligned}
\label{eqn:robertson}
\end{equation}

\noindent where $x_1$, $x_2$, and $x_3$ are the species abundances, and $p_1$, $p_2$, and $p_3$ are the reaction rates. We generated 1000 trajectories (split into training, validation, and test) with initial state $\vec{x}_{initial} = [1, 10^{-30}, 10^{-30}]$. Reaction rates $\vec{p}$ were sampled log-uniformly from the intervals $[2\times10^{-3}, 6\times10^{-3}]$, $[1.5\times10^{7}, 3.5\times10^{7}]$, and $[5\times10^{3}, 6\times10^{4}]$, spanning times $10^{-5}$ to $10^{5}$~s. The original trajectories were integrated using the Radau IIA method \citep{HairerWanner1996}. We log$_{10}$-transformed all inputs (including time), normalizing species with a z-score transformation and scaling time to $[0,1]$. This served as a baseline with which we could test different machine learning architectures on an established and similar surrogate model problem.

\begin{table}[!tbhp]
    \centering
    \footnotesize
    \setlength{\tabcolsep}{5pt}
    \begin{tabular}{ll}
        \toprule
        \textbf{Property} & \textbf{Value} \\
        \midrule

        \multicolumn{2}{l}{\textbf{MLP}} \\
        \hspace{1em} Architecture & [4, 32, 32, 32, 32, 32, 32, 32, 32, 3] \\
        \hspace{1em} Parameters & 7,651 \\
        \hspace{1em} Test Set log$_{10}$-MAE & \({1.23 \times 10^{-2}}\) \\
        \addlinespace

        \multicolumn{2}{l}{\textbf{Flow-map}} \\
        \hspace{1em} Encoder & [3, 32, 32, 16] \\
        \hspace{1em} Propagator & [17, 32, 32, 32, 32, 16] \\
        \hspace{1em} Decoder & [16, 32, 32, 3] \\
        \hspace{1em} Latent dim & \({N_{z} = 16}\) \\
        \hspace{1em} Parameters & 7,683 \\
        \hspace{1em} Test Set log$_{10}$-MAE & \({6.46 \times 10^{-3}}\) \\
        \addlinespace

        \multicolumn{2}{l}{\textbf{DeepONet}} \\
        \hspace{1em} Branch & [3, 32, 32, 32, 32, 16] \\
        \hspace{1em} Trunk & [1, 32, 32, 32, 32, 16] \\
        \hspace{1em} Features & 16 \\
        \hspace{1em} Parameters & 7,635 \\
        \hspace{1em} Test Set log$_{10}$-MAE & \({2.38 \times 10^{-2}}\) \\
        \bottomrule
    \end{tabular}
    \caption{Performance benchmarks of various architectures on the Robertson system. For this benchmark, all models take $(p_1,p_2,p_3,\Delta t)$ as inputs from the fixed initial state $\mathbf{x}(t_0)=[1,10^{-30},10^{-30}]$ and predict $\mathbf{x}(t_0+\Delta t)$. All models shared common training hyperparameters, datasets, and SiLU activation functions. Network sizes were chosen so that each model would have approximately the same number of trainable parameters.}
    \label{tab:robertson_arch_params}
\end{table}

\begin{figure}[!tbhp]
\centering
\includegraphics[width=1.0\linewidth]{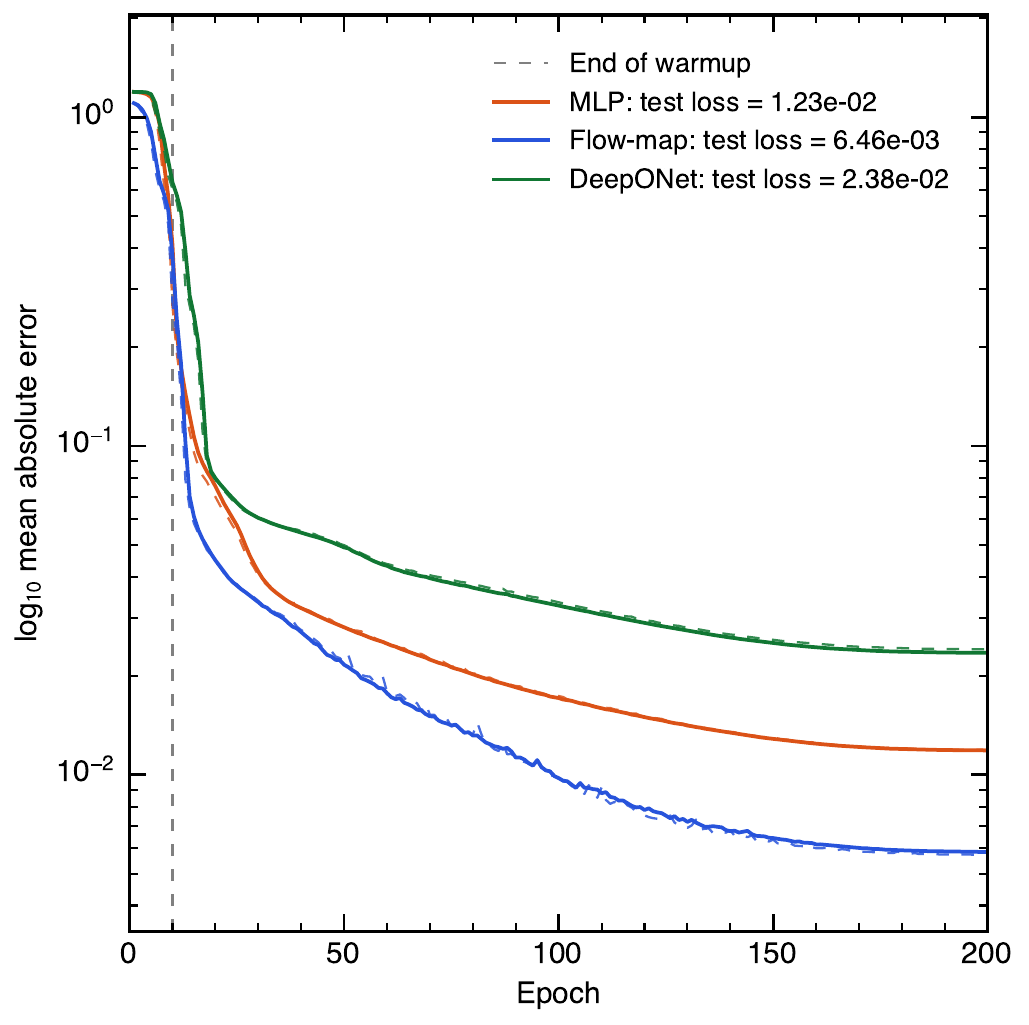}
\caption{Training curves for different model architectures on the Robertson chemical reaction system. Solid lines are training loss and dashed lines are validation loss; the legend reports the test-set log$_{10}$ mean absolute error of each architecture. The flow-map achieves the lowest test error, outperforming the DeepONet by a factor of $\sim4$ and the MLP by a factor of $\sim2$.}
\label{fig:3}
\end{figure}

Figure~\ref{fig:3} shows the training curves for a number of different model architectures  (MLP, DeepONet, and flow-map). Table~\ref{tab:robertson_arch_params} shows the architectures and final performance of each model. Overall, all model architectures are able to learn the Robertson dynamics. However, the flow-map achieves the lowest test error, roughly a factor of two lower than the MLP and a factor of $\sim4$ lower than the DeepONet. The flow-map's encoder--propagator--decoder structure imposes an information bottleneck (latent dim $N_z=16$, less than the hidden layer width of 32), so the model is not architecturally collapsible to a deeper MLP. While the DeepONet model performed worst, we note that we did not perform an exhaustive search of model hyperparameters to find the best performance. Other works \citep{Nockolds2025} suggest using separate DeepONets for each $x_i$ component (i.e., each degree of freedom in the ODE). However, we do find that this performance gap is indicative of the DeepONet model being less well suited to this specific class of vector-to-vector problem.

By changing the reaction rates, $\vec{p}$, we can alter the Robertson equations to become stiffer. When we recreate the MLP, flow-map, and DeepONet models with a new dataset with a stiffness metric of $\sim10^{16}$, the accuracy of all models was decreased. However, the results remained qualitatively consistent; the relative performance of the DeepONet, MLP, and flow-map models was unchanged.

DeepONets have been used for chemical-kinetics emulation \citep[e.g.,][]{Kumar2024, Goswami2024, Nockolds2025}. However, we note that DeepONets face a limitation for the specific chemical kinetics problem considered here. The general problem is to advance a chemical state in a single jump by some $\Delta t$. DeepONets encode the input (the system state) in the branch net and the coordinate at which the output is evaluated in the trunk net. For this problem, this trunk net coordinate represents the time evolution. However, during inference we only ever evaluate one time jump per call, and only from a single input state. For this vector-to-vector, single-query problem, one therefore loses out on many of the advantages that motivate the DeepONet framework in function-to-function problems.

\clearpage
\newpage

\bibliography{references}{}

\begin{thebibliography}{}
\expandafter\ifx\csname natexlab\endcsname\relax\def\natexlab#1{#1}\fi
\providecommand{\url}[1]{\href{#1}{#1}}
\providecommand{\dodoi}[1]{doi:~\href{http://doi.org/#1}{\nolinkurl{#1}}}
\providecommand{\doeprint}[1]{\href{http://ascl.net/#1}{\nolinkurl{http://ascl.net/#1}}}
\providecommand{\doarXiv}[1]{\href{https://arxiv.org/abs/#1}{\nolinkurl{https://arxiv.org/abs/#1}}}

\bibitem[{T. Berkemeier {et~al.}(2023)Berkemeier, Kr{\"u}ger, Feinberg,
  M{\"u}ller, P{\"o}schl, \& Krieger}]{Berkemeier2023}
Berkemeier, T., Kr{\"u}ger, M., Feinberg, A., {et~al.} 2023,
  \bibinfo{title}{Accelerating Models for Multiphase Chemical Kinetics through
  Machine Learning with Polynomial Chaos Expansion and Neural Networks,}
  Geoscientific Model Development, 16, 2037, \dodoi{10.5194/gmd-16-2037-2023}

\bibitem[{J. {Brandstetter} {et~al.}(2022){Brandstetter}, {Worrall}, \&
  {Welling}}]{Brandstetter2022}
{Brandstetter}, J., {Worrall}, D., \& {Welling}, M. 2022,
  \bibinfo{title}{{Message Passing Neural PDE Solvers},} arXiv e-prints,
  arXiv:2202.03376, \dodoi{10.48550/arXiv.2202.03376}

\bibitem[{S.~L. {Brunton} {et~al.}(2021){Brunton}, {Budi{\v{s}}i{\'c}},
  {Kaiser}, \& {Kutz}}]{Brunton2021}
{Brunton}, S.~L., {Budi{\v{s}}i{\'c}}, M., {Kaiser}, E., \& {Kutz}, J.~N. 2021,
  \bibinfo{title}{{Modern Koopman Theory for Dynamical Systems},} arXiv
  e-prints, arXiv:2102.12086, \dodoi{10.48550/arXiv.2102.12086}

\bibitem[{S.~L. Brunton \& J.~N. Kutz(2022)Brunton \& Kutz}]{Brunton2022}
Brunton, S.~L., \& Kutz, J.~N. 2022, Data-Driven Science and Engineering:
  Machine Learning, Dynamical Systems, and Control, 2nd edn. (Cambridge
  University Press)

\bibitem[{K. {Champion} {et~al.}(2019){Champion}, {Lusch}, {Kutz}, \&
  {Brunton}}]{Champion2019}
{Champion}, K., {Lusch}, B., {Kutz}, J.~N., \& {Brunton}, S.~L. 2019,
  \bibinfo{title}{{Data-driven discovery of coordinates and governing
  equations},} Proceedings of the National Academy of Science, 116, 22445,
  \dodoi{10.1073/pnas.1906995116}

\bibitem[{R.~T. Chen {et~al.}(2018)Chen, Rubanova, Bettencourt, \&
  Duvenaud}]{Chen2018a}
Chen, R.~T., Rubanova, Y., Bettencourt, J., \& Duvenaud, D.~K. 2018,
  \bibinfo{title}{Neural ordinary differential equations,} Advances in neural
  information processing systems, 31

\bibitem[{V. {Churchill} \& D. {Xiu}(2023){Churchill} \& {Xiu}}]{Churchill2023}
{Churchill}, V., \& {Xiu}, D. 2023, \bibinfo{title}{{Flow Map Learning for
  Unknown Dynamical Systems: Overview, Implementation, and Benchmarks},} arXiv
  e-prints, arXiv:2307.11013, \dodoi{10.48550/arXiv.2307.11013}

\bibitem[{W. {Cole Nockolds} {et~al.}(2025){Cole Nockolds}, {Krishnanunni},
  {Bui-Thanh}, \& {Tang}}]{Nockolds2025}
{Cole Nockolds}, W., {Krishnanunni}, C.~G., {Bui-Thanh}, T., \& {Tang}, X.
  2025, \bibinfo{title}{{Lilan: A linear latent network approach for real-time
  solutions of stiff, nonlinear, ordinary differential equations},} arXiv
  e-prints, arXiv:2501.08423, \dodoi{10.48550/arXiv.2501.08423}

\bibitem[{C.~S. Cooper \& A.~P. Showman(2006)Cooper \& Showman}]{Cooper2006}
Cooper, C.~S., \& Showman, A.~P. 2006, \bibinfo{title}{Dynamics and
  Disequilibrium Carbon Chemistry in Hot Jupiter Atmospheres, with Application
  to HD 209458b,} \apj, 649, 1048, \dodoi{10.1086/506312}

\bibitem[{P. Cvitanovi{\'c} {et~al.}(2016)Cvitanovi{\'c}, Artuso, Mainieri,
  Tanner, \& Vattay}]{ChaosBook}
Cvitanovi{\'c}, P., Artuso, R., Mainieri, R., Tanner, G., \& Vattay, G. 2016,
  Chaos: Classical and Quantum (Copenhagen: Niels Bohr Inst.).
\newblock \url{http://ChaosBook.org/}

\bibitem[{B. {Drummond} {et~al.}(2018){Drummond}, {Mayne}, {Manners}, {Carter},
  {Boutle}, {Baraffe}, {H{\'e}brard}, {Tremblin}, {Sing}, {Amundsen}, \&
  {Acreman}}]{Drummond2018b}
{Drummond}, B., {Mayne}, N.~J., {Manners}, J., {et~al.} 2018,
  \bibinfo{title}{Observable Signatures of Wind-driven Chemistry with a Fully
  Consistent Three-dimensional Radiative Hydrodynamics Model of HD 209458b,}
  \apjl, 855, L31, \dodoi{10.3847/2041-8213/aab209}

\bibitem[{B. {Drummond} {et~al.}(2020){Drummond}, {H{\'e}brard}, {Mayne},
  {Venot}, {Ridgway}, {Changeat}, {Tsai}, {Manners}, {Tremblin}, {Abraham},
  {Sing}, \& {Kohary}}]{Drummond2020}
{Drummond}, B., {H{\'e}brard}, E., {Mayne}, N.~J., {et~al.} 2020,
  \bibinfo{title}{{Implications of three-dimensional chemical transport in hot
  Jupiter atmospheres: Results from a consistently coupled
  chemistry-radiation-hydrodynamics model},} \aap, 636, A68,
  \dodoi{10.1051/0004-6361/201937153}

\bibitem[{S. {Goswami} {et~al.}(2024){Goswami}, {Jagtap}, {Babaee}, {Susi}, \&
  {Karniadakis}}]{Goswami2024}
{Goswami}, S., {Jagtap}, A.~D., {Babaee}, H., {Susi}, B.~T., \& {Karniadakis},
  G.~E. 2024, \bibinfo{title}{{Learning stiff chemical kinetics using extended
  deep neural operators},} Computer Methods in Applied Mechanics and
  Engineering, 419, 116674, \dodoi{10.1016/j.cma.2023.116674}

\bibitem[{P. {Goyal} {et~al.}(2017){Goyal}, {Doll{\'a}r}, {Girshick},
  {Noordhuis}, {Wesolowski}, {Kyrola}, {Tulloch}, {Jia}, \& {He}}]{Goyal2017}
{Goyal}, P., {Doll{\'a}r}, P., {Girshick}, R., {et~al.} 2017,
  \bibinfo{title}{{Accurate, Large Minibatch SGD: Training ImageNet in 1
  Hour},} arXiv e-prints, arXiv:1706.02677, \dodoi{10.48550/arXiv.1706.02677}

\bibitem[{E. Hairer \& G. Wanner(1996)Hairer \& Wanner}]{HairerWanner1996}
Hairer, E., \& Wanner, G. 1996, Springer Series in Computational Mathematics,
  Vol.~14, Solving Ordinary Differential Equations II: Stiff and
  Differential-Algebraic Problems, 2nd edn. (Berlin: Springer)

\bibitem[{K. {He} {et~al.}(2016){He}, {Zhang}, {Ren}, \& {Sun}}]{He2016}
{He}, K., {Zhang}, X., {Ren}, S., \& {Sun}, J. 2016, \bibinfo{title}{{Deep
  Residual Learning for Image Recognition},} in 2016 IEEE Conference on
  Computer Vision and Pattern Recognition (CVPR), 1,
  \dodoi{10.1109/CVPR.2016.90}

\bibitem[{J.~L.~A.~M. {Hendrix} {et~al.}(2023){Hendrix}, {Louca}, \&
  {Miguel}}]{Hendrix2023}
{Hendrix}, J. L.~A.~M., {Louca}, A.~J., \& {Miguel}, Y. 2023,
  \bibinfo{title}{{Using a neural network approach to accelerate disequilibrium
  chemistry calculations in exoplanet atmospheres},} MNRAS, 524, 643,
  \dodoi{10.1093/mnras/stad1763}

\bibitem[{K. Heng(2017)Heng}]{Heng2017}
Heng, K. 2017, Exoplanetary Atmospheres: Theoretical Concepts and Foundations
  (Princeton, NJ: Princeton University Press), \dodoi{10.1515/9781400883073}

\bibitem[{G.~E. Hinton \& R.~R. Salakhutdinov(2006)Hinton \&
  Salakhutdinov}]{Hinton2006}
Hinton, G.~E., \& Salakhutdinov, R.~R. 2006, \bibinfo{title}{Reducing the
  Dimensionality of Data with Neural Networks,} Science, 313, 504,
  \dodoi{10.1126/science.1127647}

\bibitem[{M.~W. Hirsch {et~al.}(2013)Hirsch, Smale, \&
  Devaney}]{HirschSmaleDevaney2013}
Hirsch, M.~W., Smale, S., \& Devaney, R.~L. 2013, Differential Equations,
  Dynamical Systems, and an Introduction to Chaos, 3rd edn. (Academic Press),
  \dodoi{10.1016/C2009-0-61160-0}

\bibitem[{S. Hochreiter \& J. Schmidhuber(1997)Hochreiter \&
  Schmidhuber}]{Hochreiter1997}
Hochreiter, S., \& Schmidhuber, J. 1997, \bibinfo{title}{Long Short-Term
  Memory,} Neural Computation, 9, 1735, \dodoi{10.1162/neco.1997.9.8.1735}

\bibitem[{K. Hornik {et~al.}(1989)Hornik, Stinchcombe, \& White}]{Hornik1989}
Hornik, K., Stinchcombe, M., \& White, H. 1989, \bibinfo{title}{Multilayer
  feedforward networks are universal approximators,} Neural Networks, 2, 359,
  \dodoi{https://doi.org/10.1016/0893-6080(89)90020-8}

\bibitem[{R. {Hu} {et~al.}(2012){Hu}, {Seager}, \& {Bains}}]{hu2012a}
{Hu}, R., {Seager}, S., \& {Bains}, W. 2012, \bibinfo{title}{{Photochemistry in
  Terrestrial Exoplanet Atmospheres. I. Photochemistry Model and Benchmark
  Cases},} ApJ, 761, 166, \dodoi{10.1088/0004-637X/761/2/166}

\bibitem[{Y. Huang \& J.~H. Seinfeld(2022)Huang \& Seinfeld}]{Huang2022}
Huang, Y., \& Seinfeld, J.~H. 2022, \bibinfo{title}{A Neural Network-Assisted
  Euler Integrator for Stiff Kinetics in Atmospheric Chemistry,} Environmental
  Science \& Technology, 56, 4676, \dodoi{10.1021/acs.est.1c07648}

\bibitem[{N.~S. {Keskar} {et~al.}(2016){Keskar}, {Mudigere}, {Nocedal},
  {Smelyanskiy}, \& {Tang}}]{Keskar2016}
{Keskar}, N.~S., {Mudigere}, D., {Nocedal}, J., {Smelyanskiy}, M., \& {Tang},
  P. T.~P. 2016, \bibinfo{title}{{On Large-Batch Training for Deep Learning:
  Generalization Gap and Sharp Minima},} arXiv e-prints, arXiv:1609.04836,
  \dodoi{10.48550/arXiv.1609.04836}

\bibitem[{A. Kumar \& T. Echekki(2024)Kumar \& Echekki}]{Kumar2024}
Kumar, A., \& Echekki, T. 2024, \bibinfo{title}{Combustion chemistry
  acceleration with DeepONets,} Fuel, 365, 131212

\bibitem[{A. {Lamb} {et~al.}(2016){Lamb}, {Goyal}, {Zhang}, {Zhang},
  {Courville}, \& {Bengio}}]{Lamb2016}
{Lamb}, A., {Goyal}, A., {Zhang}, Y., {et~al.} 2016, \bibinfo{title}{{Professor
  Forcing: A New Algorithm for Training Recurrent Networks},} arXiv e-prints,
  arXiv:1610.09038, \dodoi{10.48550/arXiv.1610.09038}

\bibitem[{E.~K.~H. Lee {et~al.}(2023)Lee, Tsai, Hammond, \& Tan}]{Lee2023}
Lee, E. K.~H., Tsai, S.-M., Hammond, M., \& Tan, X. 2023, \bibinfo{title}{A
  mini-chemical scheme with net reactions for 3D general circulation models.
  II. 3D thermochemical modelling of WASP-39b and HD 189733b,} \aap, 672, A110,
  \dodoi{10.1051/0004-6361/202245473}

\bibitem[{Z. {Li} {et~al.}(2020){Li}, {Kovachki}, {Azizzadenesheli}, {Liu},
  {Bhattacharya}, {Stuart}, \& {Anandkumar}}]{Li2020}
{Li}, Z., {Kovachki}, N., {Azizzadenesheli}, K., {et~al.} 2020,
  \bibinfo{title}{{Fourier Neural Operator for Parametric Partial Differential
  Equations},} arXiv e-prints, arXiv:2010.08895,
  \dodoi{10.48550/arXiv.2010.08895}

\bibitem[{I. {Loshchilov} \& F. {Hutter}(2017){Loshchilov} \&
  {Hutter}}]{Loshchilov2017}
{Loshchilov}, I., \& {Hutter}, F. 2017, \bibinfo{title}{{Decoupled Weight Decay
  Regularization},} arXiv e-prints, arXiv:1711.05101,
  \dodoi{10.48550/arXiv.1711.05101}

\bibitem[{L. {Lu} {et~al.}(2019){Lu}, {Jin}, \& {Karniadakis}}]{Lu2019}
{Lu}, L., {Jin}, P., \& {Karniadakis}, G.~E. 2019, \bibinfo{title}{{DeepONet:
  Learning nonlinear operators for identifying differential equations based on
  the universal approximation theorem of operators},} arXiv e-prints,
  arXiv:1910.03193, \dodoi{10.48550/arXiv.1910.03193}

\bibitem[{B. {Lusch} {et~al.}(2018){Lusch}, {Kutz}, \& {Brunton}}]{Lusch2018}
{Lusch}, B., {Kutz}, J.~N., \& {Brunton}, S.~L. 2018, \bibinfo{title}{{Deep
  learning for universal linear embeddings of nonlinear dynamics},} Nature
  Communications, 9, 4950, \dodoi{10.1038/s41467-018-07210-0}

\bibitem[{J.~M. {Mendon{\c{c}}a} {et~al.}(2018){Mendon{\c{c}}a}, {Tsai},
  {Malik}, {Grimm}, \& {Heng}}]{Mendonca2018}
{Mendon{\c{c}}a}, J.~M., {Tsai}, S.-m., {Malik}, M., {Grimm}, S.~L., \& {Heng},
  K. 2018, \bibinfo{title}{{Three-dimensional Circulation Driving Chemical
  Disequilibrium in WASP-43b},} ApJ, 869, 107, \dodoi{10.3847/1538-4357/aaed23}

\bibitem[{J.~I. {Moses} {et~al.}(2011){Moses}, {Visscher}, {Fortney},
  {Showman}, {Lewis}, {Griffith}, {Klippenstein}, {Shabram}, {Friedson},
  {Marley}, \& {Freedman}}]{moses2011}
{Moses}, J.~I., {Visscher}, C., {Fortney}, J.~J., {et~al.} 2011,
  \bibinfo{title}{{Disequilibrium Carbon, Oxygen, and Nitrogen Chemistry in the
  Atmospheres of HD 189733b and HD 209458b},} ApJ, 737, 15,
  \dodoi{10.1088/0004-637X/737/1/15}

\bibitem[{A. {Paszke} {et~al.}(2019){Paszke}, {Gross}, {Massa}, {Lerer},
  {Bradbury}, {Chanan}, {Killeen}, {Lin}, {Gimelshein}, {Antiga}, {Desmaison},
  {K{\"o}pf}, {Yang}, {DeVito}, {Raison}, {Tejani}, {Chilamkurthy}, {Steiner},
  {Fang}, {Bai}, \& {Chintala}}]{Paszke2019}
{Paszke}, A., {Gross}, S., {Massa}, F., {et~al.} 2019,
  \bibinfo{title}{{PyTorch: An Imperative Style, High-Performance Deep Learning
  Library},} arXiv e-prints, arXiv:1912.01703,
  \dodoi{10.48550/arXiv.1912.01703}

\bibitem[{T. Qin {et~al.}(2019)Qin, Wu, \& Xiu}]{QIN2019}
Qin, T., Wu, K., \& Xiu, D. 2019, \bibinfo{title}{Data driven governing
  equations approximation using deep neural networks,} Journal of Computational
  Physics, 395, 620, \dodoi{https://doi.org/10.1016/j.jcp.2019.06.042}

\bibitem[{C.~E. {Rasmussen} \& C.~K.~I. {Williams}(2006){Rasmussen} \&
  {Williams}}]{Rasmussen2006}
{Rasmussen}, C.~E., \& {Williams}, C. K.~I. 2006, {Gaussian Processes for
  Machine Learning} (The MIT Press)

\bibitem[{E. {Rauscher} \& K. {Menou}(2010){Rauscher} \&
  {Menou}}]{Rauscher2010}
{Rauscher}, E., \& {Menou}, K. 2010, \bibinfo{title}{{Three-dimensional
  Modeling of Hot Jupiter Atmospheric Flows},} ApJ, 714, 1334,
  \dodoi{10.1088/0004-637X/714/2/1334}

\bibitem[{H.~H. Robertson(1967)Robertson}]{Robertson1967}
Robertson, H.~H. 1967, \bibinfo{title}{The Solution of a Set of Reaction Rate
  Equations,} in Numerical Analysis: An Introduction, ed. J.~Walsh (Cambridge,
  Massachusetts: Academic Press), 178--182

\bibitem[{O. Ronneberger {et~al.}(2015)Ronneberger, Fischer, \&
  Brox}]{Ronneberger2015}
Ronneberger, O., Fischer, P., \& Brox, T. 2015, \bibinfo{title}{U-Net:
  Convolutional Networks for Biomedical Image Segmentation,} in Lecture Notes
  in Computer Science, Vol. 9351, Medical Image Computing and Computer-Assisted
  Intervention -- MICCAI 2015 (Springer International Publishing), 234--241,
  \dodoi{10.1007/978-3-319-24574-4_28}

\bibitem[{D.~E. Rumelhart {et~al.}(1986)Rumelhart, Hinton, \&
  Williams}]{Rumel1986}
Rumelhart, D.~E., Hinton, G.~E., \& Williams, R.~J. 1986,
  \bibinfo{title}{Learning Internal Representations by Error Propagation,} in
  Parallel Distributed Processing: Explorations in the Microstructure of
  Cognition, Volume 1: Foundations, ed. D.~E. Rumelhart, J.~L. McClelland, \&
  the PDP Research~Group (Cambridge, MA, USA: MIT Press), 318--362.
\newblock \url{http://dl.acm.org/citation.cfm?id=104279.104293}

\bibitem[{A.~P. {Showman} {et~al.}(2009){Showman}, {Fortney}, {Lian}, {Marley},
  {Freedman}, {Knutson}, \& {Charbonneau}}]{Showman2009}
{Showman}, A.~P., {Fortney}, J.~J., {Lian}, Y., {et~al.} 2009,
  \bibinfo{title}{{Atmospheric Circulation of Hot Jupiters: Coupled
  Radiative-Dynamical General Circulation Model Simulations of HD 189733b and
  HD 209458b},} ApJ, 699, 564, \dodoi{10.1088/0004-637X/699/1/564}

\bibitem[{M.~E. {Steinrueck} {et~al.}(2023){Steinrueck}, {Koskinen}, {Lavvas},
  {Parmentier}, {Zieba}, {Tan}, {Zhang}, \& {Kreidberg}}]{Steinrueck2023}
{Steinrueck}, M.~E., {Koskinen}, T., {Lavvas}, P., {et~al.} 2023,
  \bibinfo{title}{{Photochemical Hazes Dramatically Alter Temperature Structure
  and Atmospheric Circulation in 3D Simulations of Hot Jupiters},} ApJ, 951,
  117, \dodoi{10.3847/1538-4357/acd4bb}

\bibitem[{S.-M. Tsai {et~al.}(2018)Tsai, Kitzmann, Lyons, Mendon{\c c}a, Grimm,
  \& Heng}]{Tsai2018}
Tsai, S.-M., Kitzmann, D., Lyons, J.~R., {et~al.} 2018, \bibinfo{title}{Toward
  Consistent Modeling of Atmospheric Chemistry and Dynamics in Exoplanets:
  Validation and Generalization of the Chemical Relaxation Method,} \apj, 862,
  31, \dodoi{10.3847/1538-4357/aac834}

\bibitem[{S.-M. Tsai {et~al.}(2022)Tsai, Lee, \& Pierrehumbert}]{Tsai2022}
Tsai, S.-M., Lee, E. K.~H., \& Pierrehumbert, R. 2022, \bibinfo{title}{A
  mini-chemical scheme with net reactions for 3D general circulation models. I.
  Thermochemical kinetics,} \aap, 664, A82, \dodoi{10.1051/0004-6361/202142816}

\bibitem[{S.-M. {Tsai} {et~al.}(2017){Tsai}, {Lyons}, {Grosheintz}, {Rimmer},
  {Kitzmann}, \& {Heng}}]{Tsai2017}
{Tsai}, S.-M., {Lyons}, J.~R., {Grosheintz}, L., {et~al.} 2017,
  \bibinfo{title}{{VULCAN: An Open-source, Validated Chemical Kinetics Python
  Code for Exoplanetary Atmospheres},} ApJS, 228, 20,
  \dodoi{10.3847/1538-4365/228/2/20}

\bibitem[{S.-M. {Tsai} {et~al.}(2021){Tsai}, {Malik}, {Kitzmann}, {Lyons},
  {Fateev}, {Lee}, \& {Heng}}]{Tsai2021}
{Tsai}, S.-M., {Malik}, M., {Kitzmann}, D., {et~al.} 2021, \bibinfo{title}{{A
  Comparative Study of Atmospheric Chemistry with VULCAN},} ApJ, 923, 264,
  \dodoi{10.3847/1538-4357/ac29bc}

\bibitem[{O. Venot {et~al.}(2019)Venot, Bounaceur, Dobrijevic, H{\'e}brard,
  Cavali{\'e}, Tremblin, Drummond, \& Charnay}]{Venot2019}
Venot, O., Bounaceur, R., Dobrijevic, M., {et~al.} 2019, \bibinfo{title}{A
  reduced chemical scheme for modelling warm to hot hydrogen-dominated
  atmospheres,} \aap, 624, A58, \dodoi{10.1051/0004-6361/201834861}

\bibitem[{A. Vojtekova {et~al.}(2025)Vojtekova, Waldmann, Yip, Merín,
  Al-Refaie, \& Venot}]{Vojtekova2025}
Vojtekova, A., Waldmann, I., Yip, K.~H., {et~al.} 2025,
  \bibinfo{title}{CHEXANET: a novel approach to fast-tracking disequilibrium
  chemistry calculations for exoplanets using neural networks,} \mnras, 538,
  1690, \dodoi{10.1093/mnras/staf297}

\bibitem[{Y.~L. {Yung} {et~al.}(1984){Yung}, {Allen}, \& {Pinto}}]{yung1984}
{Yung}, Y.~L., {Allen}, M., \& {Pinto}, J.~P. 1984,
  \bibinfo{title}{{Photochemistry of the atmosphere of Titan - Comparison
  between model and observations},} ApJS, 55, 465

\bibitem[{K.~J. {Zahnle} \& M.~S. {Marley}(2014){Zahnle} \&
  {Marley}}]{Zahnle2014}
{Zahnle}, K.~J., \& {Marley}, M.~S. 2014, \bibinfo{title}{{Methane, Carbon
  Monoxide, and Ammonia in Brown Dwarfs and Self-Luminous Giant Planets},} ApJ,
  797, 41, \dodoi{10.1088/0004-637X/797/1/41}

\end{thebibliography}
\bibliographystyle{aasjournalv7}

\end{document}